\documentclass[aps,twocolumn,superscriptaddress,longbibliography]{revtex4-1}

\usepackage{amsmath,amssymb,latexsym}
\usepackage{graphicx}
\usepackage{hyperref}
\usepackage{natbib}
\setcitestyle{authoryear,round}
\usepackage{mathrsfs}
\usepackage{amscd}
\usepackage{color}
\usepackage{braket}
\DeclareSymbolFont{symbols}{OMS}{cmsy}{m}{n}
\DeclareSymbolFont{largesymbols}{OMX}{cmex}{m}{n}
\renewcommand{\bm}[1]{\boldsymbol #1}

\begin{document}

\title{
Higgs and Nambu-Goldstone modes in condensed matter physics
}

\author{Naoto Tsuji}
\affiliation{Department of Physics, University of Tokyo, Hongo, Tokyo 113-0033, Japan}
\affiliation{RIKEN Center for Emergent Matter Science (CEMS), Wako, Saitama 351-0198, Japan}
\author{Ippei Danshita}
\affiliation{Department of Physics, Kindai University, Higashi-Osaka, Osaka 577-8520, Japan}
\author{Shunji Tsuchiya}
\affiliation{Department of Physics, Chuo University, Kasuga, Tokyo 112-8551, Japan}

\begin{abstract}
Collective dynamics of many particle systems is tightly linked to their underlying symmetry and phase transitions.
Higgs and Nambu-Goldstone modes are, respectively, collective amplitude and phase modes of the order parameter
that are widely observed in various physical systems at different energy scales,
ranging from magnets, superfluids, superconductors to our universe. 
The Higgs mode is a massive excitation, which is a condensed-matter analog of Higgs particle in high-energy physics, 
while the Nambu-Goldstone mode is a massless excitation that appears when a continuous
symmetry is spontaneously broken.
They provide important information on the fundamental aspects of many particle systems, such as symmetry, phases, 
dynamics, response to external fields, and so on.
In this article, we review the physics of Higgs and Nambu-Goldstone modes in condensed matter physics.
Especially, we focus on the development on the study of collective modes in superconductors and cold-atom systems.
\end{abstract}


\date{\today}

\maketitle

\section{Key objectives}
\begin{itemize}
    \item Introduce collective modes of an order parameter in condensed matter physics, including Higgs and Nambu-Goldstone modes.
    \item Describe basic theories for Higgs and Nambu-Goldstone modes in superconductors and cold-atom systems (including fermionic and bosonic superfluids).
    \item Review experimental observations of Higgs and Nambu-Goldstone modes in condensed matter systems.
\end{itemize}

\section{Introduction}
A phenomenon of phase transition is widely observed in various physical systems at different energy scales,
ranging from condensed-matter to high-energy physics. For example, when temperature decreases,
a paramagnet turns into a ferromagnet, and a metal turns into a superconductor.
Those phenomena can be universally understood with the concept of spontaneous symmetry breaking: 
Physical laws are invariant under a certain symmetry transformation, while a physically realized state at low energy does not necessarily exhibit the same symmetry.
In the case of magnets having spin rotation symmetry, 
individual spins are randomly oriented at high temperature, whereas at low temperature they are aligned to
one particular direction which is spontaneously chosen. In the case of superconductors, a complex phase 
of a macroscopic wavefunction of electrons is aligned spontaneously.

\begin{figure}[b]
\includegraphics[width=8cm]{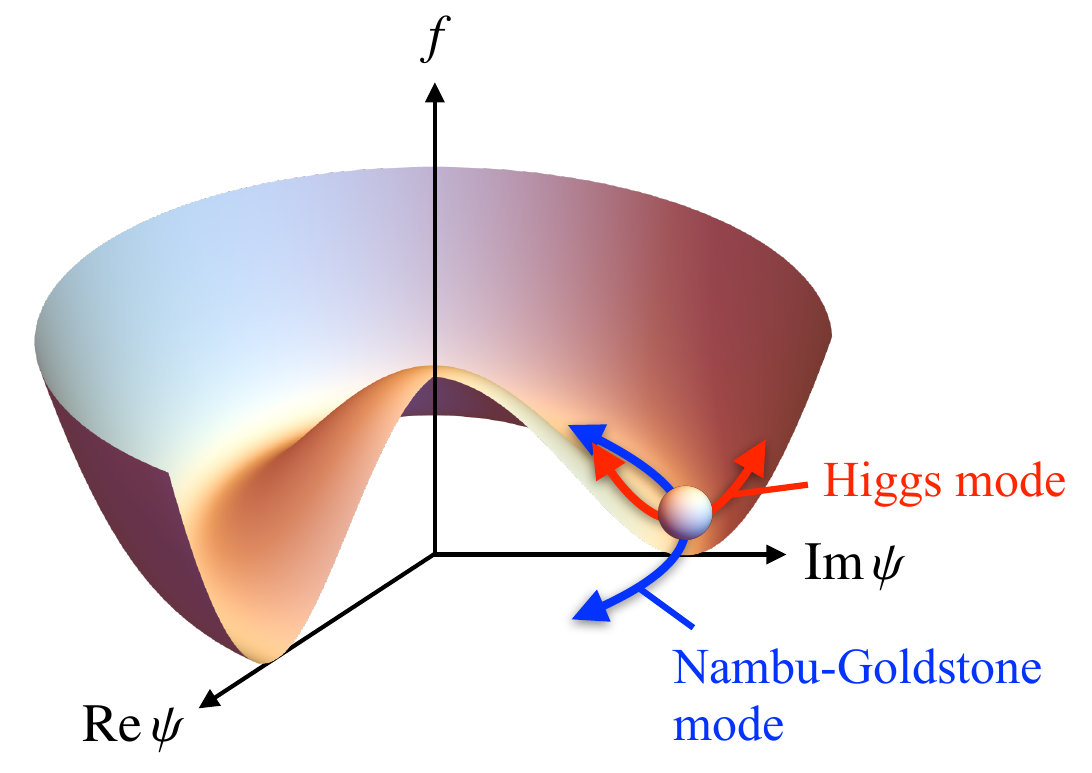}
\caption{Free-energy surface in the plane of a complex order parameter $\psi$. The red (blue) arrows represent
Higgs (Nambu-Goldstone) mode, corresponding to amplitude (phase) fluctuation of the order parameter.}
\label{mexican hat}
\end{figure}

In the conventional Landau paradigm, phase transition is characterized by an order parameter, which
grows in ordered phases, such as magnetization or condensate (superfluid) density. When a continuous symmetry is spontaneously broken,
there typically appear two types of collective motions of the order parameter: 
One is a phase fluctuation of the order parameter,
and the other is an amplitude fluctuation of the order parameter (Fig.~\ref{mexican hat}). 
The former is often referred to as Nambu-Goldstone (NG) mode,
while the latter is called Higgs mode, which recently attracts interests in condensed matter physics.
As the name suggests, the Higgs mode is a condensed-matter analog of a Higgs particle in high-energy physics,
which emerges as a result of quantization of the amplitude fluctuation of a complex scalar field.
The Higgs particle plays an important role in generating a mass of gauge fields due to the Higgs mechanism,
making a force mediated by gauge bosons short-ranged.

Historically, on the other hand, the physics of Higgs particle and Higgs mechanism arose from the study of superconductivity. 
The macroscopic phenomenological theory of superconductivity had been developed by Ginzburg and Landau in 1950
(\cite{GinzburgLandau1950}),
which can be viewed as a non-relativistic version of the complex scalar field theory adopted by Higgs and others
in the study of the Higgs mechanism. The microscopic theory of superconductivity had been presented by Bardeen, Cooper, and Schrieffer in 1957 (\cite{BCS1957}), known as the BCS theory, which had a broad impact on various branches of physics. In 1958, just one year after the BCS theory,
Anderson and Bogoliubov had already discussed collective modes in superconductors (\cite{BogoliubovBook, Anderson1958, Anderson1958b}). Especially, Anderson mentioned
the existence of what is known as the Higgs mode in superconductors, 
whose energy was found to be twice as large as the superconducting gap.
Physically, the Higgs mode in superconductors corresponds to a collective oscillation of the superfluid density (density of condensed 
Cooper pairs), which should be distinguished from an individual excitation of Cooper-pair breaking.

\begin{table*}[t]
\caption{Examples of phase and amplitude modes in various physical systems.
In the case of superconductors and the standard model in high-energy physics, 
the system is coupled to gauge fields, and the phase mode disappears due to the Anderson-Higgs mechanism.}
\label{table of collective modes}
\begin{tabular}{cccc}
\hline
system & Nambu-Goldstone (phase) mode & Higgs (amplitude) mode & broken symmetry \\
\hline
\hline
magnet & magnon & amplitude mode & spin rotation \\
\hline
(incommensurate) charge density wave & phason & amplitudon & space translation\\
\hline
superfluid $^4$He
& phase mode & --- & phase rotation \\
\hline
weakly interacting superfluid Bose gas
& phase mode & --- & phase rotation \\
\hline
\begin{tabular}{c}
strongly interacting superfluid Bose gas in \\
an optical lattice at commensurate filling
\end{tabular}
& phase mode & amplitude mode & phase rotation \\
\hline
superfluid Fermi gas & phase mode & amplitude mode & phase rotation \\
\hline
superfluid $^3$He (B phase) & phase mode & squashing mode & SO(3)$\times$SO(3)$\times$U(1) \\
\hline
superconductor & --- & amplitude mode & phase rotation \\
\hline
atomic nuclei & pion & $\sigma$ particle & chiral symmetry \\
\hline
standard model in high-energy physics & --- & Higgs particle & SU(2)$\times$U(1) \\
\hline
\end{tabular}
\end{table*}

Stimulated by the development of the BCS theory, Nambu had introduced the concept of spontaneous symmetry breaking in
particle physics. Based on an analogy with the BCS theory (\cite{Nambu1960}), a possibility of spontaneous symmetry breaking is postulated
for a vacuum ground state in a relativistic field theory (\cite{Nambu1961, Goldstone1961}). It was found that a massless boson appears
as a phase fluctuation of the order parameter in a symmetry broken phase,
in much the same way as a spin wave appears as a collective excitation in magnets.
In parallel, Goldstone and others proved a theorem (\cite{Goldstone1962}) that massless particles must appear 
when a continuous symmetry is spontaneously broken in a relativistic field theory, 
and the number of those correspond to the number of degrees of broken symmetries. 
Later, the theorem has been extended to non-relativistic cases (\cite{Watanabe2012}, \cite{Hidaka2013}), where the counting rule of the NG modes has been established.

In the application of symmetry breaking in particle physics, one has to avoid Goldstone's theorem, since no such massless particle had
been observed in our universe. In 1964, Englert, Brout, Higgs, and others had noticed that there is a way to go around Goldstone's theorem
when the system is coupled to gauge fields (\cite{Englert1964, Higgs1964}, \cite{Guralnik1964}): The phase mode is absorbed 
into the longitudinal component of gauge fields, which results in generating a mass of gauge bosons (Higgs mechanism).
What remains after this is the amplitude mode of the scalar field, which corresponds to
a massive scalar particle (Higgs boson). In 2012, the Higgs particle has been finally discovered in the LHC experiment at CERN (\cite{ATLAS2012, CMS2012}),
after almost 50 years since the theoretical prediction.

It should be emphasized that the Higgs mechanism had been discussed earlier by Anderson 
in the context of superconductivity in condensed matter physics (\cite{Anderson1958, Anderson1958b, Anderson1963}): In superconductors, 
an interior magnetic field is completely excluded due to the Meissner-Ochsenfeld effect.
This phenomenon can be understood by the fact that electromagnetic fields acquire a mass and cannot propagate freely
inside superconductors (Anderson-Higgs mechanism). Anderson's argument was based on a non-relativistic theoretical formalism,
but essentially the same thing happens in relativistic field theories.

\begin{figure}[t]
\includegraphics[width=7cm]{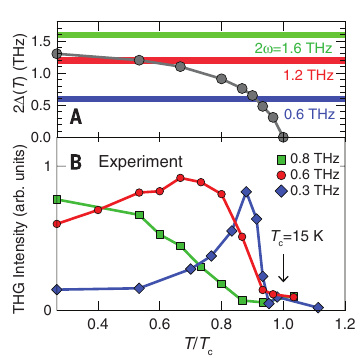}
\caption{Temperature dependence of (A) the superconducting gap compared with twice the pump frequencies and (B) the intensity of third harmonic generation in NbN superconductors. (Adapted from \cite{Matsunaga2014})}
\label{fig: THG}
\end{figure}

While the physics of Higgs phenomena has originated from the study of superconductivity, 
the Higgs mode in superconductors (\cite{PekkerVarma2015, ShimanoTsuji2020}), which manifests the existence of an effective scalar potential
with a mexican-hat shape (Fig.~\ref{mexican hat}) that causes the Higgs mechanism,
has not been observed for a long time. An exception at an early stage was the Raman scattering experiment
reported in 1980 for 2H-NbSe$_2$ (\cite{SooryakumarKlein1980}),
which is rather a special material in the sense that superconductivity coexists with a charge-density-wave (CDW) order.
Only in this particular regime, a resonance peak corresponding to the amplitude mode has been observed in Raman spectra.

In 2013, a coherent long-lived oscillation induced 
by a mono-cycle terahertz laser pulse
was observed for a pure superconductor NbN 
(`pure' means the one without coexistence of any other long-range orders) (\cite{Matsunaga2013}).
The oscillation frequency agrees with the superconducting gap energy, implying that 
the Higgs mode is excited in the terahertz laser experiment.
In the subsequent experiment, a multi-cycle terahertz pulse has been applied to NbN, 
showing a coherent oscillation with the frequency twice as large as the pump laser frequency
(\cite{Matsunaga2014}).
This is consistent with the fact that the Higgs mode can be generated by a two-photon absorption process through the nonlinear light-Higgs coupling (\cite{Tsuji2015}). At the same time, the resonant enhancement of third harmonic generation has been observed in NbN when the superconducting gap coincides with twice the pump frequency (Fig.~\ref{fig: THG}) (\cite{Matsunaga2014}). Later, it turns out that 
impurities play an essential role in the resonant enhancement of third harmonic generation mediated by the Higgs mode (\cite{Jujo2018, Murotani2019, Silaev2019, Tsuji2020}, \cite{Seibold2021}).

If one is not limited to charged systems in which the order parameter is coupled with gauge fields,
the phase and amplitude modes of order parameters
have been widely observed in symmetry broken phases (see Table~\ref{table of collective modes}).
For example, 
magnons, or spin waves, correspond to
phase modes in magnetically ordered systems. 
Liquid helium and cold atomic gases are another examples of physical systems
in which collective modes of bosonic and fermionic superfluids are well studied.
These collective modes provide important information on the underlying system,
such as symmetries, phases, dynamics, response to external fields, and so on.

Throughout the article, we put $\hbar=1$.

\section{Higgs mode in superconductors}

The emergence of collective modes in symmetry broken phases can be understood with a simple macroscopic phenomenological model.
For example, the low-energy effective theory of superconductors corresponds to the Ginzburg-Landau (GL) theory, whose Lagrangian density is given,
as a functional of a complex scalar field $\psi(\bm r)$ (a `macroscopic wavefunction' of electrons), by
\begin{align}
\mathcal L
&=
-\left[ 
a|\psi(\bm r)|^2+\frac{b}{2}|\psi(\bm r)|^4
+\frac{1}{2m^\ast}|(-i\nabla-e^\ast\bm A)\psi(\bm r)|^2\right]
\notag
\\
&\quad
+c_1 \psi(\bm r)^\dagger (i\partial_t-e^\ast\phi)\psi(\bm r)
+c_2 |(i\partial_t-e^\ast\phi)\psi(\bm r)|^2.
\label{GL Lagrangian}
\end{align}
Here $a, b, c_1$ and $c_2$ are constants, $m^\ast$ and $e^\ast$ are effective mass and charge of condensates,
and $\phi$ and $\bm A$ are scalar and vector potentials for external electromagnetic fields, respectively. 
The time derivative terms are included phenomenologically.
The amplitude of the scalar field $\psi(\bm r)$ serves as an order parameter, and its squared absolute value
has a meaning of the superfluid density ($|\psi(\bm r)|^2=n_s$).
The Lagrangian is invariant under the global phase rotation, $\psi\to e^{i\theta}\psi$.
The coefficient $a$ is assumed to behave as $a=a_0(T-T_c)$ $(a_0>0)$ near the critical point,
where $T$ is a temperature of the system, and $T_c$ is a critical temperature at which superconductivity emerges.
At high temperature, the effective potential (the first two terms in the square bracket in Eq.~(\ref{GL Lagrangian})) has a parabolic cylindrical shape with a single minimum at the origin ($\psi=0$).
As the temperature goes down below $T_c$, the potential turns into a mexican-hat shape (Fig.~\ref{mexican hat}),
where degenerate minima appear along the circle at the bottom of the potential ($|\psi|=\psi_0>0$). 
One of the minima is realized as the ground state, which spontaneously breaks the phase rotation symmetry.
Without loss of generality, one can assume that the order parameter takes a real positive value in the ground state.

One can consider fluctuations of the order parameter around the ground state by expanding as
\begin{align}
\psi(\bm r)
&=
(\psi_0+H(\bm r))e^{i\theta(\bm r)},
\end{align}
where $H(\bm r)$ and $\theta(\bm r)$ represent amplitude and phase fluctuations, respectively.
The Lagrangian is then expanded in the following way:
\begin{align}
\mathcal L
&=
2aH^2-\frac{1}{2m^\ast}(\nabla H)^2
-\frac{e^{\ast 2}}{2m^\ast}\left(\bm A-\frac{1}{e^\ast}\nabla\theta\right)^2
\notag
\\
&\quad
\times (\psi_0+H)^2
-c_1 e^\ast\left(\phi+\frac{1}{e^\ast}\partial_t\theta\right)
(\psi_0+H)^2
\notag
\\
&\quad
+c_2 (\partial_t H)^2
+c_2 e^{\ast 2}\left(\phi+\frac{1}{e^\ast}\partial_t\theta\right)^2
(\psi_0+H)^2
\notag
\\
&\quad
+\cdots.
\label{GL expanded}
\end{align}
The first term indicates that the Higgs mode represented by $H$ has a finite mass, reflecting the fact that
the effective potential has a finite curvature along the amplitude direction.
The second term in Eq.~(\ref{GL expanded}) corresponds to the kinetic term of the Higgs mode.
The phase field $\theta$ appears without a mass term ($\propto \theta^2$).
Hence, $\theta$ would be thought of as a massless Nambu-Goldstone mode.
However, $\theta$ always appears in the combination of $\bm A-\frac{1}{e^\ast}\nabla\theta$ or $\phi+\frac{1}{e^\ast}\partial_t\theta$,
due to which one can remove $\theta$ by performing a gauge transformation,
$\bm A\to \bm A'=\bm A-\frac{1}{e^\ast}\nabla\theta$ and $\phi\to \phi'=\phi+\frac{1}{e^\ast}\partial_t\theta$ (unitary gauge).
By rewriting $\bm A'$ and $\phi'$ as $\bm A$ and $\phi$, one obtains an expression in terms of $H$,
\begin{align}
\mathcal L
&=
2aH^2-\frac{1}{2m^\ast}(\nabla H)^2-\frac{e^{\ast 2}}{2m^\ast}\bm A^2 (\psi_0+H)^2
\notag
\\
&\quad
-c_1 e^\ast \phi (\psi_0+H)^2
+c_2(\partial_t H)^2
+c_2 e^{\ast 2} \phi^2(\psi_0+H)^2
\notag
\\
&\quad
+\cdots.
\label{GL expanded 2}
\end{align}
The phase mode $\theta$ disappears from the expression, and a mass term of the gauge field ($\propto \bm A^2$) appears in the third term.
This is nothing but the Anderson-Higgs mechanism.
One can count the number of degrees of freedom before and after the Anderson-Higgs mechanism:
\begin{align}
&
2\; (\mbox{transverse components of }\bm A)
\notag
\\
&\!\!
+2\; (\mbox{real and imaginary parts of $\psi$})
\notag
\\
&\!\!\to
3\; (\mbox{transverse and longitudinal components of }\bm A)
\notag
\\
&\quad
+1\; (\mbox{Higgs mode})
\notag
\end{align}

The electromagnetic field, that mediates the long-range Coulomb interaction,
screens phase fluctuations of the superconducting order parameter, pushing them up to high energy
(in the scale of plasma frequency). As a result of the Anderson-Higgs mechanism,
the electromagnetic field itself becomes short-ranged, and decays exponentially within the London penetration depth
(Meissner-Ochsenfeld effect). The remaining low-energy excitation mode is the Higgs mode.
In the vicinity of $T_c$, however, the phase mode may have a possibility to revive without damping (known as the Carlson-Goldman mode (\cite{CarlsonGoldman1975}))
due to the presence of a large number of normal electrons which screen the long-range Coulomb interaction.

If the system is electrically neutral (i.e., $e^\ast=0$), the order parameter is not coupled to gauge fields,
and the Anderson-Higgs mechanism does not take place. In this case, the NG mode survives as a massless excitation,
which dominates the low-energy behavior. In particular, the Higgs mode can hybridize with the NG mode if $c_1\neq 0$,
which causes a mixing between amplitude and phase modes.
In order to decouple the amplitude and phase modes, one often requires an effective particle-hole symmetry, which
suppresses the $c_1$ term and makes the Lagrangian relativistic with an emergent Lorentz symmetry.

In superconductors, there is an approximate particle-hole symmetry at low energy, so that one can practically neglect the $c_1$ term.
From Eq.~(\ref{GL expanded 2}), one can see that there is no linear coupling between $H$ and $\bm A$ (or $\phi$).
This means that the Higgs mode cannot be excited within the linear response regime, which is well expected
from the fact that the Higgs mode is a scalar excitation having no electric charge or magnetic moment.
On the other hand, if one goes beyond the linear response, the Higgs mode can be excited by electromagnetic fields
through the nonlinear coupling (e.g., $\propto \bm A^2 H$) (\cite{Tsuji2015}), which has been used to observe the Higgs mode
in terahertz laser experiments.

The Ginzburg-Landau theory does not predict the precise value of the mass of the Higgs mode.
It only tells that the mass is proportional to $(-a)^{\frac{1}{2}}\propto (T_c-T)^{\frac{1}{2}}$.
From a microscopic theory, the mass is found to coincide with the superconducting gap energy $2\Delta$.
That is, the minimum energy that is required to excite the Higgs mode is the same as the energy
to break a single Cooper pair. This causes a problem of 
using the Ginzburg-Landau theory in the energy scale of the Higgs mode: In the GL theory,
the order parameter is supposed to vary sufficiently slowly in space and time,
whereas the order parameter oscillates with the frequency of $2\Delta$ when the Higgs mode is excited.
This already reaches the energy scale of pair breaking, and thus one cannot neglect the effect of 
quasiparticle excitations, invalidating the basic assumption of the GL description.

The aforementioned issue motivates one to move to a microscopic description of superconductivity,
i.e., the BCS theory. In the BCS theory, one adopts the Hamiltonian that includes the kinetic energy of electrons
and the pairing interaction (usually mediated by phonons),
\begin{align}
H_{\rm BCS}
&=
\sum_{\bm k,\sigma} \varepsilon_{\bm k} c_{\bm k\sigma}^\dagger c_{\bm k\sigma}
-\frac{V}{N}\sum_{\bm k, \bm k'} c_{\bm k\uparrow}^\dagger c_{-\bm k\downarrow}^\dagger
c_{-\bm k'\downarrow}c_{\bm k'\uparrow},
\label{BCS Hamiltonian}
\end{align}
where $c_{\bm k\sigma}$ is the annihilation operator of electrons with momentum $\bm k$ and spin $\sigma$,
$\varepsilon_{\bm k}$ is the energy dispersion, $V (>0)$ is the pairing interaction strength,
and $N$ is the number of $\bm k$ points.

In the mean-field approximation, the Hamiltonian (\ref{BCS Hamiltonian}) is replaced by
$H_{\rm MF}=\sum_{\bm k} \Psi_{\bm k}^\dagger h(\bm k)\Psi_{\bm k}$, where
$\Psi_{\bm k}=(c_{\bm k\uparrow}\; c_{-\bm k\downarrow}^\dagger)^T$ is the Nambu spinor and
\begin{align}
h(\bm k)
&=
\begin{pmatrix}
\varepsilon_{\bm k} & -\Delta^\ast \\
-\Delta & -\varepsilon_{\bm k}
\end{pmatrix}
\end{align}
is the Bogoliubov-de Gennes Hamiltonian.
The off-diagonal element $\Delta$ is defined by
\begin{align}
\Delta
&=
\frac{V}{N}
\sum_{\bm k} \langle c_{\bm k\uparrow}^\dagger c_{-\bm k\downarrow}^\dagger \rangle.
\end{align}
We assume that $\Delta$ takes a real positive value.
The eigenvalues of $h(\bm k)$ is found to be $\pm E_{\bm k}:=\pm\sqrt{\varepsilon_{\bm k}^2+\Delta^2}$, and thus the single-particle spectrum shows
an energy gap of $2\Delta$. At zero temperature, $\Delta$ satisfies the gap equation,
\begin{align}
\Delta
&=
\frac{V}{N}
\sum_{\bm k} \frac{\Delta}{2E_{\bm k}}.
\label{gap equation}
\end{align}

The dynamics of superconductors is determined by the time-dependent Bogoliubov-de Gennes equation,
\begin{align}
i\frac{\partial}{\partial t}\Psi_{\bm k}
&=
h(\bm k)\Psi_{\bm k}.
\label{BdG equation}
\end{align}
To analyze the equation,
it is convenient to introduce Anderson's pseudospins,
\begin{align}
\sigma_{\bm k}^\alpha
&=
\frac{1}{2}
\langle\Psi_{\bm k}^\dagger \tau^\alpha \Psi_{\bm k}\rangle
\quad
(\alpha=x,y,z).
\end{align}
Here $\tau^\alpha$ denote the Pauli matrices. Physically, 
the $x$ and $y$ components of the pseudospins correspond, respectively, to the real and imaginary parts
of Cooper pairs' amplitude $\langle c_{\bm k\uparrow}^\dagger c_{-\bm k\downarrow}^\dagger\rangle$, 
and the $z$ component represents the momentum-dependent occupation of electrons.
From Eq.~(\ref{BdG equation}), one finds that the time evolution of the pseudospins is given by
\begin{align}
\frac{\partial}{\partial t}\bm \sigma_{\bm k}
&=
2\bm b_{\bm k}\times\bm \sigma_{\bm k},
\label{Bloch equation}
\end{align}
which resembles the Bloch equation that describes precession motion of spins in a magnetic field.
Here $\bm b_{\bm k}=(-{\rm Re}\,\Delta, -{\rm Im}\,\Delta, \varepsilon_{\bm k})$ plays a role of a magnetic field
for pseudospins.

In general, the equation of motion (\ref{Bloch equation}) becomes a nonlinear equation,
since it is supplemented by the self-consistency condition (\ref{gap equation}).
However, one can linearize the equation if the deviation from the initial ground state is sufficiently small.
By writing $\bm \sigma_{\bm k}(t)=\bm\sigma_{\bm k}(0)+\delta\bm\sigma_{\bm k}(t)$
and $\Delta(t)=\Delta+\delta\Delta(t)$,
the linearized equation is expressed as
\begin{align}
\frac{\partial}{\partial t}\delta\sigma_{\bm k}^x
&=
-2\varepsilon_{\bm k} \delta\sigma_{\bm k}^y,
\\
\frac{\partial}{\partial t}\delta\sigma_{\bm k}^y
&=
2\varepsilon_{\bm k} \delta\sigma_{\bm k}^x+2\Delta \delta\sigma_{\bm k}^z-\frac{\varepsilon_{\bm k}}{E_{\bm k}}\delta\Delta,
\\
\frac{\partial}{\partial t}\delta\sigma_{\bm k}^z
&=
-2\Delta\delta\sigma_{\bm k}^y.
\end{align}
By using the relation $\Delta\delta\sigma_{\bm k}^x=\varepsilon_{\bm k}\delta\sigma_{\bm k}^z$,
one can further simplify the equation into
\begin{align}
\frac{\partial^2}{\partial t^2}\delta\sigma_{\bm k}^x
&=
-4E_{\bm k}^2\delta\sigma_{\bm k}^x+\frac{2\varepsilon_{\bm k}^2}{E_{\bm k}}\delta\Delta.
\end{align}
After Fourier transformation, one obtains 
\begin{align}
\delta\sigma_{\bm k}^x(\omega)
&=
\frac{2\varepsilon_{\bm k}^2}{E_{\bm k}(4E_{\bm k}^2-\omega^2)}\delta\Delta(\omega)
\end{align}
Combining with the linearized gap equation, $\delta\Delta(\omega)=\frac{V}{N}\sum_{\bm k} \delta\sigma_{\bm k}^x(\omega)$,
one finds that a nontrivial solution exists if the following condition is satisfied:
\begin{align}
\frac{V}{N}\sum_{\bm k}\frac{2\varepsilon_{\bm k}^2}{E_{\bm k}(4E_{\bm k}^2-\omega^2)}
&=
1.
\label{Higgs mode equation}
\end{align}
This is precisely the condition that has been derived by Anderson in 1958 (\cite{Anderson1958}). One can quickly verify
that $\omega=2\Delta$ is a solution of Eq.~(\ref{Higgs mode equation}), 
since it is reduced to the gap equation (\ref{gap equation}) at this frequency.
The solution $\Delta(t)$ consists of a collective amplitude oscillation
of the order parameter, and hence is identified to the Higgs mode that has been expected to exist in the GL theory.

Subsequently, the equation of motion for a quench problem has been studied by Volkov and Kogan in 1974 (\cite{VolkovKogan1973}), who showed that the amplitude oscillation
of the gap function does not persist but decays in a power law as
\begin{align}
\Delta(t)
\sim
\Delta+\frac{c}{\sqrt{2\Delta t}}\cos(2\Delta t+\varphi)
\label{eq.volkov}
\end{align}
with certain constants $c$ and $\varphi$. This slight instability (power-law decay) of the Higgs mode is related to the fact that
the energy of the Higgs mode lies right at the bottom edge of the quasiparticle excitation continuum.
The decay into quasiparticles occurs without collisions,
similar to the case of Landau damping for plasma waves. 
The energy dispersion of the Higgs mode at finite momentum, $\omega_H(\bm q)$, has 
the following form (\cite{LittlewoodVarma1982}),
\begin{align}
\omega_H(\bm q)
&=
\sqrt{(2\Delta)^2+\frac{1}{3}v_F^2 q^2}-i\frac{\pi^2}{24}v_F q,
\end{align}
where $v_F$ is the Fermi velocity and $q=|\bm q|$. The presence of the imaginary part indicates that the Higgs mode
has a finite lifetime at finite $\bm q$ due to the decay into quasiparticles.
Especially, the mode becomes overdamped when $v_F q/\Delta \gtrsim 1$.

In multiband superconductors, the system may possess multiple order parameters (multi-gap superconductors) such as MgB$_2$.
In this case, there can arise a collective oscillation of the relative phase between two order parameters, known as the Leggett mode (\cite{Leggett1966}).
Since the relative phase oscillation does not induce the net current and hence does not couple to gauge fields linearly,
the Leggett mode can evade the Anderson-Higgs mechanism, 
remaining to survive at low energy.

If there is a hidden competing order that stays close to the primary order, 
the so-called Bardasis-Schrieffer mode can emerge as a collective oscillation of the amplitude of the secondary
order parameter (\cite{BardasisSchrieffer1961}). 
For example, some of iron-based superconductors are expected to have a close competition 
between the $s_\pm$ pairing ground state with the $d$-wave pairing instability. 
The Bardasis-Schrieffer mode can be used to probe such hidden instabilities that are difficult to detect by other means.

In strongly correlated systems, one may need to go beyond the BCS theory and take into account quantum correction effects to treat Higgs modes, which can be defined by the peak of the superconducting amplitude-amplitude correlation function. It is expected that the lifetime (i.e., inverse of the peak width) of the Higgs mode would become shorter and shorter as one goes into strongly correlated regimes due to the decay to quasiparticles, and the amplitude oscillation of the order parameter would become overdamped at some point. Such kind of behavior has been observed in the ordered phase of the Hubbard model with an interaction quench (\cite{Werner2012, Tsuji2013}).

\section{Higgs modes in atomic Fermi gases}

The Higgs mode not only appears in superconductors but also in neutral fermionic superfluids. The only traditional condensed-matter system that falls into the latter category had been the system of superfluid $^3{\rm He}$. In the last few decades, a new system arose in the field of atomic physics: Superfluidity in a dilute gas of fermionic atoms has been realized in 2004 (\cite{Regal2004},\cite{Zwierlein2004}). 
Since then, this system has been offering a unique opportunity for studying Higgs modes in fermionic superfluids.

Atomic gases have several advantages over solid state systems. One of them is high controllability of system parameters. Using experimental techniques developed in atomic, molecular, and optical physics, various parameters can be tuned with unprecedented precision. In atomic Fermi gases, the interaction strength between fermions can be tuned by external fields via the Feshbach resonance. Employing this technique, one can observe a smooth crossover between the BCS-type state with large overlapping Cooper pairs and Bose-Einstein condensation (BEC) of tightly bound molecules of two fermions, known as the ``BEC-BCS crossover", which has been realized also in 2004 (\cite{Regal2004},\cite{Zwierlein2004}). This phenomenon had never been directly observed in condensed-matter systems, though it was theoretically speculated by Leggett (\cite{Leggett1980}) and Eagles (\cite{Eagles1969}). This experimental progress allows one to study the evolution of the Higgs mode in the BEC-BCS crossover.

The interaction strength in this system is often described by the dimensionless parameter $1/(k_F a)$, where $k_F$ and $a$ are the Fermi wave number and the $s$-wave scattering length, respectively. The $s$-wave scattering length can be tuned by an external magnetic field due to the Feshbach resonance. The weak-coupling BCS regime and strong-coupling BEC regime correspond to $1/(k_Fa)<0$ and $1/(k_Fa)>0$, respectively. The crossover between the BCS and BEC physics occurs at the unitary limit $1/(k_Fa)=0$, at which $a$ diverges.

\begin{figure}[t]
\includegraphics[width=7cm]{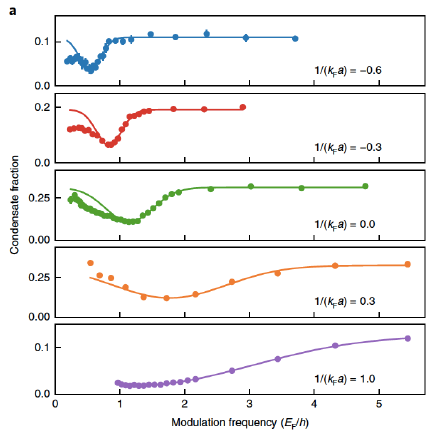}
\caption{Measured excitation spectra of the Higgs mode in a superfluid Fermi gas for different interaction strength $1/(k_Fa)$. (Adapted from \cite{Behrle2018})}
\label{fig:spectrum}
\end{figure}

The Higgs mode in a superfluid Fermi gas of $^6$Li atoms has been observed in 2018 (\cite{Behrle2018}). A novel scheme to induce the Higgs mode has been developed in the experiment: One of the two hyperfine states involved in pairing is coupled with an initially unoccupied third state by a radiofrequency field. In this manner, a periodic modulation of the amplitude of the gap function was induced.
The Higgs mode was searched by rapidly sweeping the magnetic field onto the molecular side of the Feshbach resonance and measuring the energy absorption spectrum. The excitation spectra in Fig.~\ref{fig:spectrum} show the clear resonance at $2\Delta$ in the BCS side ($1/(k_Fa)<0$), while the peak broadens significantly signalling the instability of the Higgs mode in the BEC side ($1/(k_Fa)>0$).
\par
The time-dependent Bogoliubov-de Gennes (tdBdG) formalism has been extensively used to study the dynamical properties of superfluid Fermi gases. In this formalism, the gap function $\Delta(\bm r,t)$ obeys the tdBdG equation 
\begin{align}
i\frac{\partial}{\partial t}
\begin{pmatrix}
u_\nu(\bm r,t)\\
v_\nu(\bm r,t)
\end{pmatrix}
=
\begin{pmatrix}
\hat h & \Delta \\
\Delta^* & -\hat h
\end{pmatrix}
\begin{pmatrix}
u_\nu(\bm r,t)\\
v_\nu(\bm r,t)
\end{pmatrix},
\label{eq.tdBdG}
\end{align}
where $\hat h=-\nabla^2/2m+U(\bm r)-\mu$, $m$ is the atomic mass, $\mu$ is the chemical potential, 
$U(\bm r)$ is the trapping potential, $\nu$ is the index for the eigenstates, 
and $u_\nu$ and $v_\nu$ are the time-dependent quasi-particle amplitudes, respectively.
The tdBdG equation needs to be solved self-consistently with the gap equation,
\begin{align}
\Delta(\bm r,t)=\frac{V}{N}\sum_\nu u_\nu(\bm r,t)v_\nu^*(\bm r,t),
\label{eq.tdBdG_gap}
\end{align}
and the equation for the total number of atoms $\mathcal N$,
\begin{align}
{\mathcal N}=2\int d\bm r\sum_\nu |u_\nu(\bm r,t)|^2.
\label{eq.tdBdG_number}
\end{align}

In a uniform system ($U(\bm r)=0$), substituting the static solution of Eq.~(\ref{eq.tdBdG}) $u_{\bm k}^2=(1+\varepsilon_{\bm k}/E_{\bm k})/2$ and $v_{\bm k}^2=(1-\varepsilon_{\bm k}/E_{\bm k})/2$ ($\varepsilon_{\bm k}=k^2/2m-\mu$ and $E_{\bm k}=\sqrt{\varepsilon_{\bm k}^2+\Delta^2}$) into Eqs.~(\ref{eq.tdBdG_number}) and (\ref{eq.tdBdG_number}), the former reduces to the gap equation (\ref{gap equation}), while the latter reduces to
\begin{equation}
\mathcal N=\sum_{\bm k}\left(1-\frac{\varepsilon_{\bm k}}{E_{\bm k}}\right).
\label{eq.number}
\end{equation}
The self-consistent solution of $(\Delta,\mu)$ for the gap equation (\ref{gap equation}) and the number equation (\ref{eq.number}) smoothly connects the weak-coupling BCS limit with $((8\epsilon_F/e^2)e^{-\pi/2k_F|a|},\epsilon_F)$
and the strong-coupling BEC limit with $(\sqrt{16/3\pi}|\mu|^{1/4}\epsilon_F^{3/4},-1/2ma^2)$,
where $e$ is Napier's constant and $\epsilon_F$ is the Fermi energy. Note that $\mu$ coincides with the binding energy of molecules in the BEC limit. The above tdBdG formalism thus correctly captures the dynamical properties of a superfluid Fermi gas including the Higgs mode throughout the entire BEC-BCS crossover. 

One can study small amplitude oscillations of the gap function (Higgs mode) induced by a quench of the interaction strength in the tdBdG formalism. The gap function evolves in time according to Eq.~(\ref{eq.volkov}) in the BCS regime, where the amplitude oscillates with the frequency $\omega_H=2\Delta$ and the oscillation decays in a power law ($\sim t^\gamma$) with the exponent $\gamma=-1/2$ (\cite{VolkovKogan1973}). 
In the BEC regime, on the other hand, the frequency $\omega_H=2\sqrt{\Delta^2+\mu^2}$ and the exponent of the power-law decay $\gamma=-3/2$ for the amplitude oscillation have been predicted (\cite{Gurarie2009}). The former coincides with twice of the energy threshold for the creation of fermionic excitations by pair-breaking. The evolution of the frequency $\omega_H$ and the power $\gamma$ through the BEC-BCS crossover is of particular interest and worth investigating in future experiments (\cite{Tokimoto2019}). 

The NG mode has also been investigated in atomic fermi superfluids. It manifest itself as the Anderson-Bogoliubov (AB) phonon (phase) mode as shown in TABLE \ref{table of collective modes}. Anderson predicted the sound velocity $c=v_F/\sqrt{3}$ for the AB phonon mode in the BCS limit by the generalized random-phase approximation (\cite{Anderson1958}). The evolution of the excitation spectra throughout the BEC-BCS crossover has been studied using two-photon Bragg spectroscopy, which allows characterization of the dispersion relation of the AB phonon mode (\cite{Hoinka2017}).

\section{Higgs and Nambu-Goldstone modes in other systems}
The NG and Higgs modes exist ubiquitously in quantum many-body systems that possess both a spontaneously broken continuous symmetry and an approximate particle-hole symmetry. In the previous sections, we have taken superconductors and superfluid states of ultracold two-component Fermi gases as specific examples that possess either or both of the NG and Higgs modes. In this section, we explain several other significant examples of the NG and Higgs modes in the contexts of condensed matter and ultracold-atom physics in order to illustrate broad applicability of the concepts.

\begin{figure}[htbp]
\includegraphics[width=8cm]{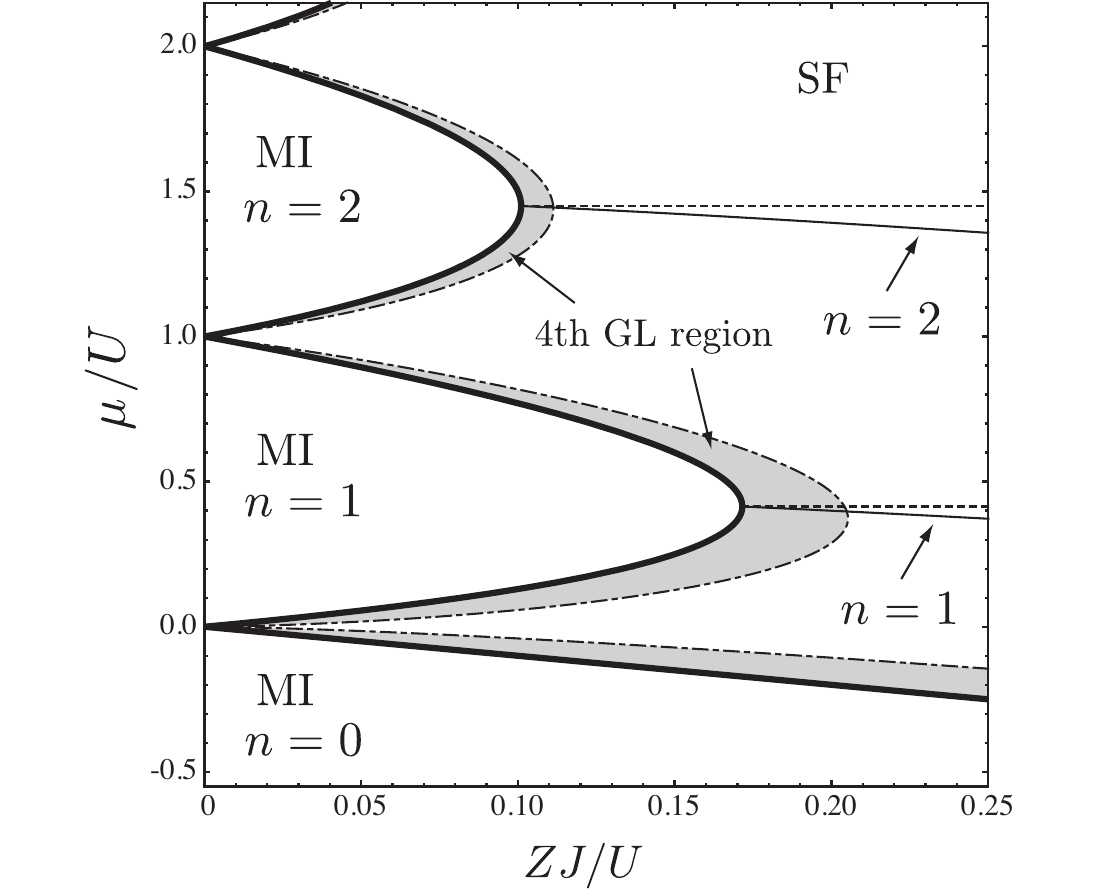}
\caption{Ground-state phase diagram of the homogeneous Bose-Hubbard model obtained by using a mean-field theory. The gray shaded areas near the Mott insulator (MI) regions roughly mark the regions where the superfluid (SF) order parameter is small enough for the fourth-order GL approximation (i.e., the effective action can be written by the expansion up to fourth order in the order parameter) to be valid. On the dashed lines in the SF region, the effective particle-hole symmetry is present. $n$ denotes the filling factor. (Adapted from \citet{Nakayama2015})
}
\label{fig:BHM_PD}
\end{figure}
We first consider a superfluid state of ultracold spinless Bose gases in optical lattices. An optical lattice means a periodic potential for atoms that is created by two counter-propagating laser beams. When the optical lattice potential is sufficiently deep, the system is well described by the Bose-Hubbard model,
\begin{eqnarray}
\hat{H}_{\rm BH} &=& - J \sum_{\langle j,l \rangle}\left(
\hat{b}_j^{\dagger}\hat{b}_l + {\rm h.c.}
\right)
+ \frac{U}{2}\sum_j \hat{b}_j^{\dagger}\hat{b}_j^{\dagger}\hat{b}_j\hat{b}_j
\nonumber \\
&& + \sum_j (\epsilon_j - \mu)\hat{b}_j^{\dagger}\hat{b}_j,
\end{eqnarray}
where $\hat{b}_j$, $U$, $J$, and $\mu$ denote, respectively, the annihilation operator of bosons at site $j$, the onsite interaction, the hopping, and the chemical potential. The external field potential $\epsilon_j$ is present for trapped atoms in standard experiments. In Fig.~\ref{fig:BHM_PD}, we show a ground-state phase diagram of the homogeneous Bose-Hubbard model in the $(ZJ/U,\mu/U)$ plane, where $Z$ is the coordination number. When the filling factor, which is the number of particles per lattice site, is integer and $ZJ/U$ is small, the ground state is a Mott insulating (MI) state. Otherwise, it is a superfluid state. There is a continuous quantum phase transition from the MI to the superfluid state, where the U(1) symmetry is spontaneously broken. In the superfluid region near the quantum phase transitions, which is the gray-shaded area in Fig.~\ref{fig:BHM_PD}, low-energy properties of the system can be effectively described by the Ginzburg-Landau (GL) theory \citep{Sachdev2011}. At the tip of the lobe of the Mott insulating region, where the quantum phase transition to a superfluid with commensurate filling occurs, there emerges particle-hole symmetry. In the nearby superfluid state at the commensurate filling, the emergent particle-hole symmetry survives approximately so that the effective theory coincides with the Lagrangian of Eq.~(\ref{GL Lagrangian}) with $e^\ast = 0$. Consequently, the Higgs mode is present as an approximately independent collective mode. Note that there is no Higgs mode in a bosonic superfluid state without particle-hole symmetry as in the cases of superfluid ${}^4$He and Bose-Einstein condensates of weakly-interacting atomic Bose gases. When $ZJ/U$ decreases towards the critical value $(ZJ/U)_{\rm c}$, the energy gap of the Higgs mode $\Delta_{\rm g}$ approaches zero as $\Delta_{\rm g} \propto |zJ/U - (zJ/U)_{\rm c}|^{\nu}$, where $\nu$ is the critical exponents for the correlation length of the $(D+1)$-dimensional classical $XY$ model and $D$ denotes the spatial dimension of the system.

\begin{figure}[t]
\includegraphics[width=8.5cm]{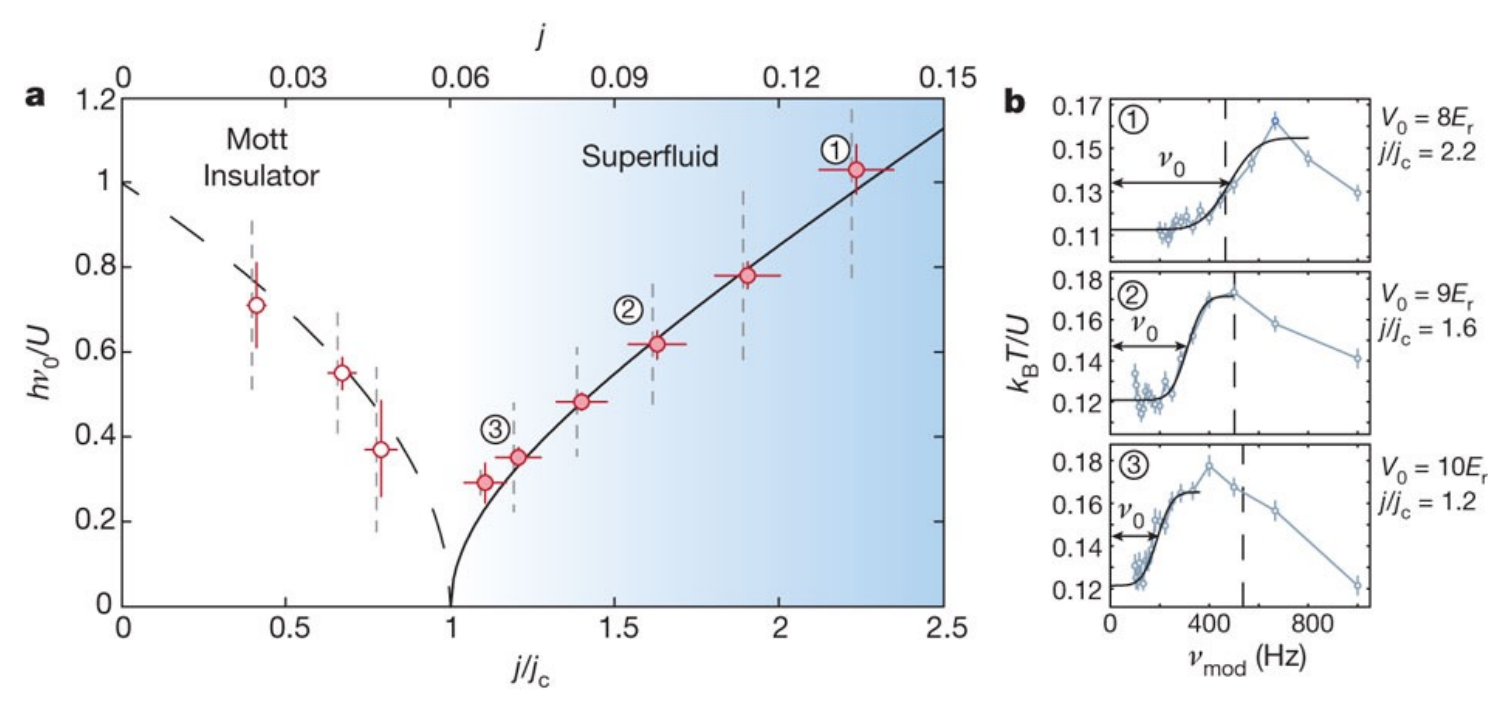}
\caption{Experimental observation of the Higgs gap in the system of ultracold Bose gases in optical lattices. {\bf a}. Excitation gap versus the hopping parameter $j$, which is denoted by $J$ in the main text, in unit of its critical value $j_{\rm c}$. Solid line and dashed line, respectively, stand for the Higgs and Mott gaps calculated by the Gutzwiller mean-field theory for a homogeneous system with unit filling.
 {\bf b}. Temperature response to temporal modulation of the lattice depth for three values of $j/j_{\rm c}$. (Adapted from \citet{Endres2012})
}
\label{fig:OLexp}
\end{figure}
In 2012, Endres {\it et al.}~have investigated dynamics of two-dimensional gas of ${}^{87}$Rb atoms (bosons) in optical lattices in response to the temporal modulation of the lattice depth in order to measure the gap of the Higgs mode (\cite{Endres2012}). Figure \ref{fig:OLexp}(b) shows the temperature of the system after the application of the lattice modulation as a function of the modulation frequency for the superfluid state at $J/J_{\rm c} = 1.2$, where $J_{\rm c}$ is the transition value of the hopping for a given $U$ at unit filling. Above an onset frequency $\nu_0$, the response exhibits a broad spectrum. Quantitative comparisons with theoretical analyses have revealed that the onset frequency (Fig.~\ref{fig:OLexp}(a)) coincides with the gap of the Higgs mode at zero momentum. It is worth noticing that the Higgs mode has not been observed as a resonance peak in the response. This smearing of the resonance can be attributed to combined effects of the spatial inhomogeneity due to the trapping potential and the finite temperatures \citep{Pollet2012}.

\begin{figure}[b]
\includegraphics[width=8cm]{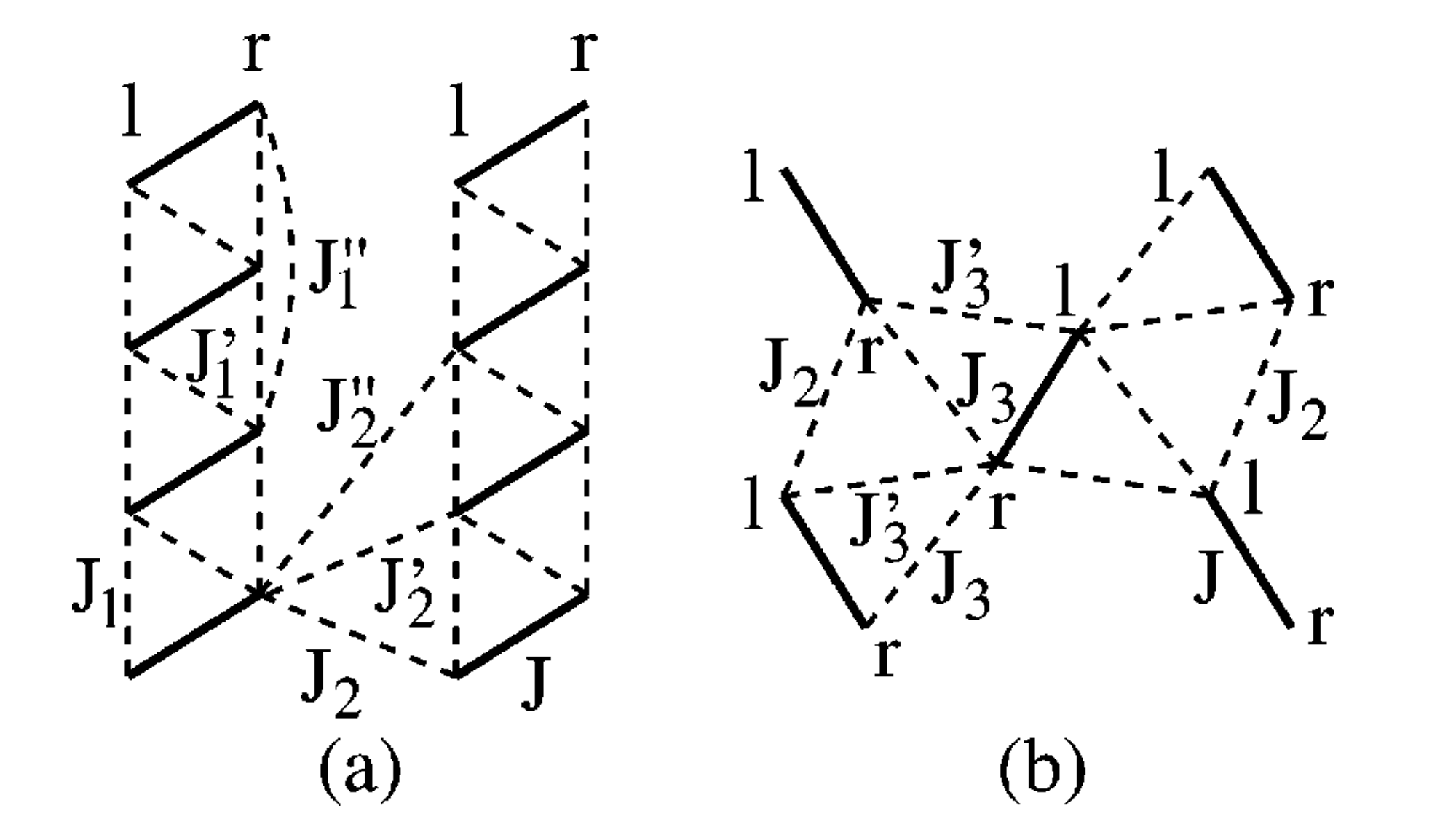}
\caption{Schematic diagram of the interdimer interaction between each pair of spins of ${\rm XCuCl}_3$ in (a) $a$-$c$ plane and (b) $b-c$ plane, where X is, e.g., Tl or K. One spin with $S=1/2$ lives at each vertex. (Adapted from \citet{Matsumoto2004})
}
\label{fig:spinEx}
\end{figure}
We next consider antiferromagnetic materials consisting of dimers of $S=1/2$ spins, such as ${\rm TlCuCl}_3$ \citep{Ruegg2008} and ${\rm KCuCl}_3$ \citep{Kuroe2012}, which can be effectively described by the following Hamiltonian,
\begin{eqnarray}
\hat{H}_{\rm AFD} &=& \sum_{j}\left(J \hat{\bf S}_{j,R}\cdot \hat{\bf S}_{j,L} - J_{xx}\hat{S}^x_{j,R}\cdot \hat{S}^x_{j,L} \right)
\nonumber \\
&&+ \sum_{j,l;m,m'}J_{j,l;m,m'} \hat{\bf S}_{l,m}\cdot \hat{\bf S}_{j,m'}.
\end{eqnarray}
Here, $\hat{\bf S}_{j,m}=(\hat{S}^x_{j,m},\hat{S}^y_{j,m},\hat{S}^z_{j,m})$ denotes the $S=1/2$ spin operator at spin $m$($=L,R$) of dimer $j$. $J(>0)$, $J_{xx}(>0)$ and $J_{j,l;m,m'}$ are the intradimer isotropic interaction, the intradimer anisotropic interaction, and the interdimer isotropic interaction, respectively. In Fig.~\ref{fig:spinEx}, the spatial configuration of the interdimer interactions is illustrated. To be concrete, we hereafter focus on the case of ${\rm TlCuCl}_3$, in which the Higgs mode has been observed via neutron scattering experiments. In this case, the most dominant interdimer interaction is $J_2$ that is antiferromagnetic. Although $J_{xx}$ is about 1\% of $J$, it is important for discussing the NG and Higgs modes in the sense that it reduces the spin rotation symmetry from SU(2) to U(1).

At atmospheric pressure, $J$ is dominant over $J_2$ so that the low-temperature magnetic state is a dimerized state in that two spins on each dimer form a spin singlet and there is no magnetic order. When the hydrostatic pressure $p$ is increased, $J_2/J$ increases and a continuous quantum phase transition to an antiferromagnetic ordered state occurs at $p=p_{\rm c}\simeq1.08$ kbar. Associated with the transition, the spin-rotation symmetry is spontaneously broken. The quantum phase transition can be qualitatively captured by a mean-field approximation in which the many-body wave function is assumed to be a product state,
\begin{eqnarray}
|\Psi_{\rm MF}\rangle = \bigotimes_j |\phi\rangle_j
\end{eqnarray}
where
\begin{eqnarray}
|\phi\rangle_j = a_{j,0}|s\rangle_j + a_{j,1}|t_1\rangle_j +  a_{j,2}|t_2\rangle_j + a_{j,3}|t_3\rangle_j.
\end{eqnarray}
The local state at dimer $j$ is spanned by the spin-singlet and triplet states, namely $\ket{s} =2^{-\frac{1}{2}}(\ket{\uparrow,\downarrow}-\ket{\downarrow,\uparrow})$, $\ket{t_1} = 2^{-\frac{1}{2}}(\ket{\uparrow,\downarrow}+\ket{\downarrow,\uparrow})$, $\ket{t_2} = 2^{-\frac{1}{2}}(\ket{\uparrow,\uparrow}+\ket{\downarrow,\downarrow})$, and $\ket{t_3} = 2^{-\frac{1}{2}}(\ket{\uparrow,\uparrow}-\ket{\downarrow,\downarrow})$. The coefficients satisfy the normalization condition $\sum_{\alpha=0}^3|a_{j,\alpha}|^2=1$.
In the limit of $J_{j,l;m,m'}\rightarrow 0$, the four states are the eigenstates of each local Hamiltonian, where $\ket{s}$ is the ground state, $\ket{t_1}$ and $\ket{t_2}$ are the degenerate first excited states, and $\ket{t_3}$ is the highest-energy state. When $J_2/J$ is finite but smaller than the critical value, the ground state remains to be $\ket{\phi}=\ket{s}$ for all the dimers and there is the energy gap to create triplet states. The energy cost for creating $\ket{t_1}$ is equivalent to that for creating $\ket{t_2}$ so that this symmetry can be regarded as a particle-hole symmetry. At the critical value of $J_2/J$, the energy gap closes. When $J_2/J$ increases further, the coefficients $a_{j,\alpha}$ for $\alpha=1,2,3$ become finite and the antiferromagnetic order emerges. The low-energy properties in the ordered phase can be effectively described by the GL theory of Eq.~(\ref{GL Lagrangian}), thanks to the spontaneous breaking of the U(1) symmetry and the presence of the particle-hole symmetry. Hence, there are the NG and Higgs modes in the ordered phase. In addition, there is another gapped mode that dominantly includes the $\ket{t_3}$ excitation.

\begin{figure}[t]
\includegraphics[width=8cm]{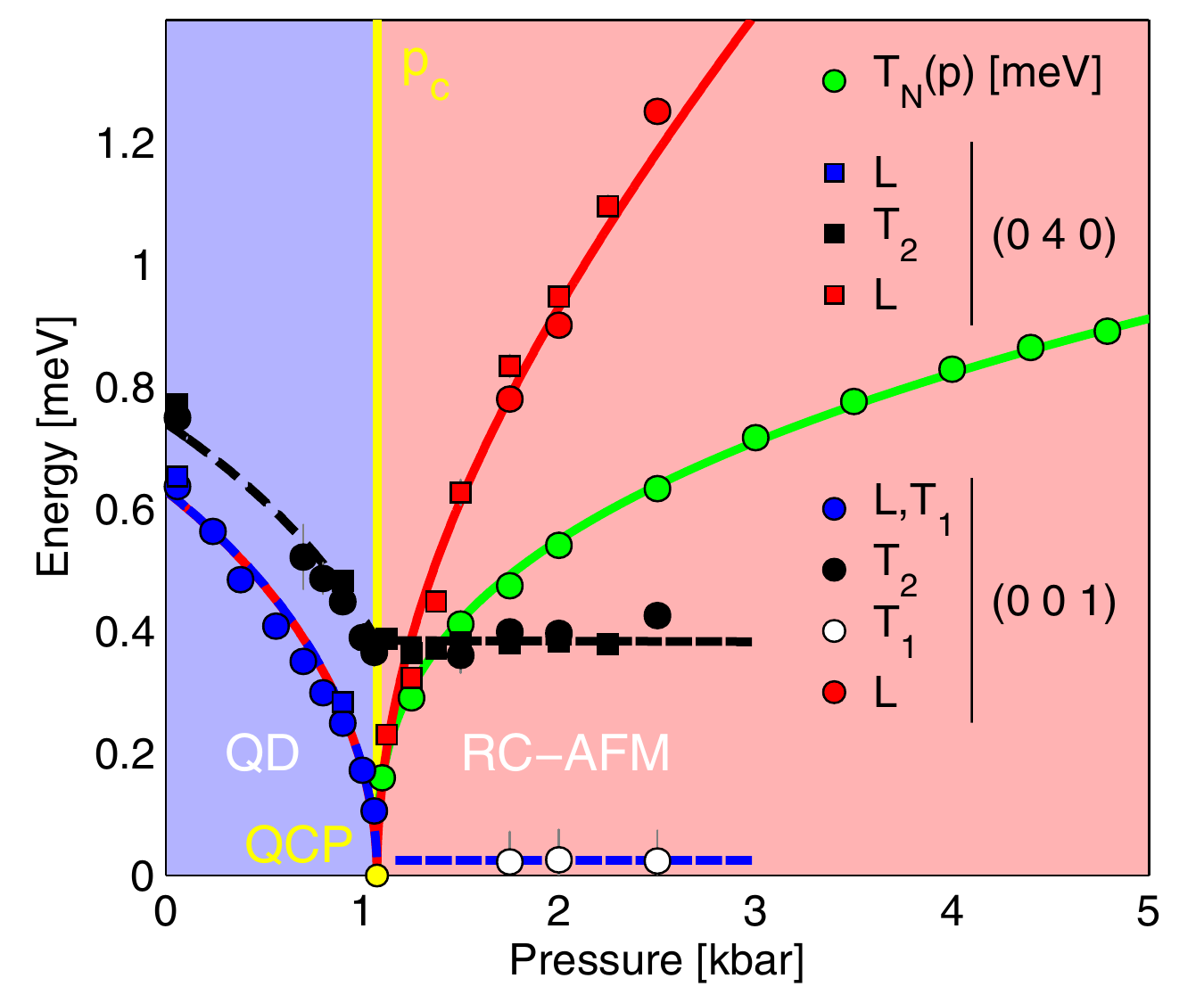}
\caption{Energy gaps of the magnetic excitations in ${\rm TlCuCl}_3$ versus pressure, which has been detected via inelastic neutron scattering. The red symbols correspond to the gap of the Higgs mode in the ordered phase, namely the renormalized classical antiferromagnetic (RC-AFM) phase. QD and QCP stand for the quantum disordered phase and quantum critical point, respectively. $p_c$, L, and T denote the critical pressure, longitudinal magnetic excitation, and transverse magnetic excitation, respectively. $(0\,4\,0)$ or $(0\,0\,1)$ means the value of the momentum transfer in the neutron scattering. (Adapted from \citet{Ruegg2008})}
\label{fig:magnetHiggs}
\end{figure}
In 2008, R\"uegg {\it et al.}~have observed the excitation modes explained above by means of inelastic neutron scattering
(\cite{Ruegg2008}). Figure~\ref{fig:magnetHiggs} shows the energy gaps of the excitation modes at specific momenta, which are detected as resonance peaks in the spectra of inelastic neutron scattering. In the dimerized phase at low pressure, there are three gapped modes corresponding to the creation of a triplet excitation. The gap of the two modes (blue symbols) closes at the quantum critical point whereas that of the other mode (black symbols) remains gappped. In the antiferromagnetic ordered phase at high pressure, the former two modes transform into the Higgs and NG modes, which correspond to longitudinal and transverse fluctuations of the magnetic order parameter.

In the last place, we briefly mention another important example, namely, materials with incommensurate CDW (ICDW) order \citep{Gruner1988}. Specifically, in solid-state materials with quasi-1D lattice structure, such as $2H$-${\rm NbSe}_2$ \citep{Tsang1976} and ${\rm K}_{0.3}{\rm MoO}_3$ \citep{Yusupov2010}, electron-phonon interactions induce the condensation of particle-hole pairs with momenta $2k_{\rm F}$ and $-2k_{\rm F}$, leading to the formation of ICDW states at low temperatures. In the ICDW state, the electron density is given by
\begin{eqnarray}
\rho(x) \simeq \rho_0 +\rho_1 \cos(2k_{\rm F}x+\varphi),
\end{eqnarray}
where $k_{\rm F}$, $\rho_0$, $\rho_1$, and $\varphi$ denote, respectively, the Fermi momentum, the average electron density, the amplitude, and the phase of ICDW. The oscillation of the amplitude $\rho_1$, which is often called amplitudon, corresponds to the Higgs mode whereas that of the phase $\varphi$ (phason) does to the NG mode. The Higgs mode has been observed in several ICDW materials via Raman spectroscopy. 

\section{Summary and outlook}

We have reviewed various order-parameter dynamics that appear in condensed matter systems. Especially, we have focused on Higgs and Nambu-Goldstone (NG) modes, corresponding to amplitude and phase oscillations of order parameters. They provide important information on the fundamental properties of underlying quantum many-body systems, including symmetries, phase transitions, dynamics, and response to external perturbations. Target systems range from superconductors to cold-atom systems (bosonic and fermionic superfluids), quantum magnets, charge density waves, and so on. Here one can see the universality of order-parameter dynamics, which may not depend so much on details of the system, partly because they are effectively described by a `common language' based on quantum field theories at low energies.

There are many open issues that remain to be addressed. In the case of superconductors, it is experimentally important to understand various aspects of Higgs modes in superconductors, especially for those beyond NbN and the coexisting phase of CDW and superconductivity in NbSe$_2$. The difficulty lies in the fact that there are always competitions between the Higgs mode and quasiparticle excitations (\cite{Cea2016}), both of which have similar characters with respect to symmetries and excitation energies. To distinguish them, one has to carefully compare experimental results with theoretical calculations, which may depend on details of materials, impurities, electron-phonon couplings, and so on (\cite{Tsuji2020}). This reflects the situation that one has not fully understood the universal (i.e., material-independent) condition that the Higgs mode becomes visible in experiments. To see enriched order-parameter dynamics, it will be fascinating to extend the observation of Higgs modes in various multi-band superconductors such as MgB$_2$ (\cite{Kovalev2021})
and unconventional superconductors (\cite{Schwarz2020}) such as cuprates (\cite{Barlas2013}, \cite{Katsumi2018}, \cite{Chu2020}) and iron-based superconductors (\cite{Vaswani2021}, \cite{Grasset2022}).
A potential application of the Higgs mode includes detection of light-induced orders (\cite{Isoyama2021, Katsumi2022}) and hidden fluctuations (\cite{Katsumi2020}).

In the case of cold-atom systems, further investigations, both experimentally and theoretically, are needed to better understand the evolution of the Higgs modes across the BCS-BEC crossover in a superfluid fermi gas and the SF-MI phase transition in a bosonic superfluid in an optial lattice. 
In the lattice-boson systems, in particular, it is a long-standing problem whether or not the Higgs mode can be detected as a resonance peak in the response to temporal modulation of an external potential. Since recent experiments on optical-lattice systems have realized almost homogeneous samples of ultracold gases (e.g., \cite{Mazurenko2017}), it will be interesting to experimentally address such a problem. 
Last but not least, taking advantage of the high degree of ability to control the system, it may be worthwhile to investigate previously unexplored aspects of Higgs modes, such as tunneling properties (\cite{Nakayama2015,Nakayama2019}), which are typically difficult to study in solid-state systems.  

\bibliography{ref}
\end{document}